\documentclass[conference]{IEEEtran}
\usepackage{array}
\usepackage{amstext}
\usepackage{graphicx}
\usepackage{epstopdf}

\makeatletter

\providecommand{\tabularnewline}{\\}

\@ifundefined{lettrine}{\usepackage{lettrine}}{}

\makeatother

\begin{document}

\title{Area Coverage Under Low Sensor Density}

\author{\IEEEauthorblockN{ Mohammad Abu Alsheikh\IEEEauthorrefmark{1}\IEEEauthorrefmark{2}, 
Shaowei Lin\IEEEauthorrefmark{2}, 
Hwee-Pink Tan\IEEEauthorrefmark{2} 
and Dusit Niyato\IEEEauthorrefmark{1}}

\IEEEauthorblockA{\IEEEauthorrefmark{1}School of Computer Engineering, Nanyang Technological University, Singapore 639798}
\IEEEauthorblockA{\IEEEauthorrefmark{2}Sense and Sense-abilities Programme, Institute for Infocomm Research, Singapore 138632}
}

\maketitle
\begin{abstract}
This paper presents a solution to the problem of monitoring a region of interest (RoI) using a set of nodes that is not sufficient to achieve the required degree of monitoring coverage. In particular, sensing coverage of wireless sensor networks (WSNs) is a crucial issue in projects due to failure of sensors. The lack of sensor equipment resources hinders the traditional method of using mobile robots to move around the RoI to collect readings. Instead, our solution employs supervised neural networks to produce the values of the uncovered locations by extracting the non-linear relation among randomly deployed sensor nodes throughout the area. Moreover, we apply a hybrid backpropagation method to accelerate the learning convergence speed to a local minimum solution. We use a real-world data set from meteorological deployment for experimental validation and analysis.
\end{abstract}

\section{Introduction}

We are interested in the problem of monitoring a region of interest
(RoI) with a limited set of nodes. In particular, this mainly occurs
when the available sensors are not enough to achieve the required
level of deployment density, e.g., coverage holes result due to temporary
node failure. Traditionally, this problem is tackled using mobile
robots that move through the uncovered points of the RoI, e.g., \cite{batalin2002sensor,costanzo2012nodes,liu2013dynamic,erdelj2013multiple}.
However, such mobile solutions are not practical in many scenarios
and suffer from many constraints. Firstly, the mobile node may not
be able to move through the area, e.g., difficulty of the terrain's
obstacles or due to human activities. Secondly, as mobile nodes move
among the designed locations, the system cannot provide the readings
for all locations at all time instances. Thirdly, the mobile node
suffers from the energy limitation. Fourthly, the development and
deployment of a mobile node can be too costly.

Related solutions exploit the spatio-temporal correlation among sensor
nodes to enhance area coverage and monitoring, e.g., \cite{gupta2008efficient,michaelides2011improved,he2012leveraging}.
These solutions are utilized to allocate the best locations to monitor
the RoI using the available nodes. In particular, the RoI is divided
into sensing zones and each zone is covered by one or more sensor
nodes while maintaining the connectivity with other nodes, i.e., they
exploit the joint coverage and connectivity problem. In contrast,
we study the case in which the system suffers from severe scarcity
of deployed nodes such that the sensing zone are not fully covered,
i.e., some zones are not covered by any node. As a result, the solution
extracts non-linear relations among zones to predict the value of
the uncovered zones.

Neural networks mimic the human brain to find non-linear patterns
in data. A supervised neural network consists of an input layer, one
or more hidden layers, and an output layer. Layers are connected to
each other using synapse weights. The backpropagation algorithm \cite{rumelhart1988learning}
provides a mechanism to fit the weights of the neural networks. In
other words, the algorithm updates network weights to determine the
connection between the input and the output data. This includes two
main phases: propagation and weight tuning phase. Initially, the propagation
phase spreads the input data forward through the network to generate
the estimated output. Then, the estimated and the actual outputs are
used to calculate the error value that is moved back through the network,
i.e., from the output layer through the hidden layers to the input
layer. Therefore, the neural network regulates itself to minimize
the difference between the actual and the predicted vectors. Resilient
propagation (Rprop) \cite{riedmiller1992rprop} is a backpropagation
variant that tunes the supervised neural network\textquoteright{}s
weights by considering only the sign change of the neural network's
cost function. Another method adapts the use of the Broyden\textendash{}Fletcher\textendash{}Goldfarb\textendash{}Shanno
(BFGS) algorithm to train the neural network \cite{ngiam2011optimization}.

Our proposed algorithm is designed to predict readings from uncovered
zones. Therefore, the algorithm increases the system coverage and
support any random deployment while minimizing the operational costs.
The system will be run in a centralized processing unit. Moreover,
we show that the learning process (both execution time and performance)
can be significantly enhanced by using Rprop for a few iterations
and then BFGS for final tuning. In particular, Rprop converges faster
than BFGS at initial iterations and with a lower computational complexity.
However, BFGS outperforms Rprop in finding more accurate local minimum.

\section{Proposed algorithm}

\begin{figure}
\begin{centering}
\includegraphics[width=0.94\columnwidth,trim=1.5cm 1cm 1cm 1cm]{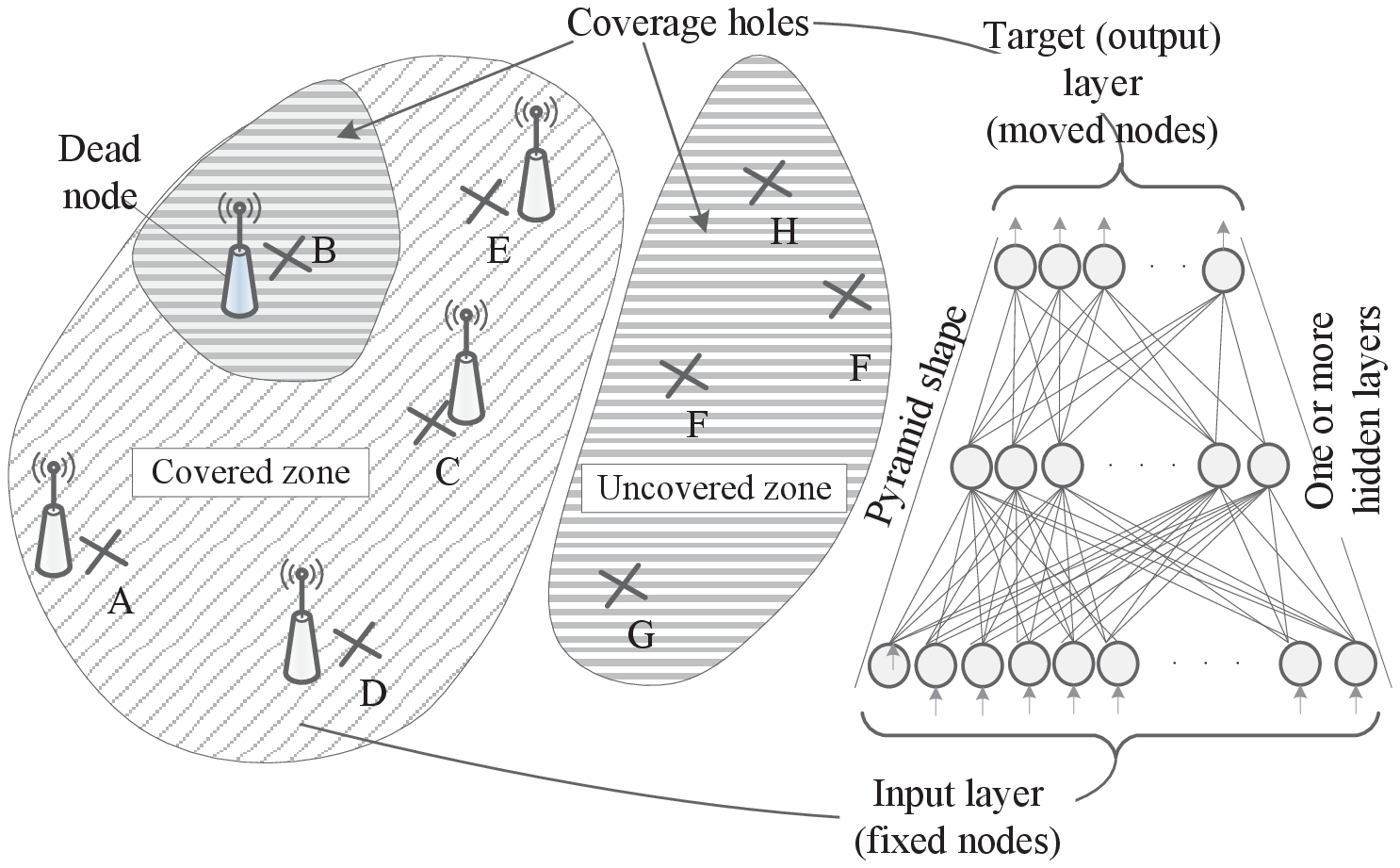}
\par\end{centering}

\caption{\label{fig:problem}The problem of monitoring the RoI with a few sensor
nodes along with a coverage hole due to a dead node.}
\end{figure}

Suppose the scenario shown in Fig. \ref{fig:problem}. The points
A to H represent the locations that the system must sense. However,
the system designer has only a few sensor nodes, e.g., due to the
limited funding, that are not sufficient to cover all monitored points.
This shortage of nodes prevents achieving the required quality of
service and results in coverage holes. The proposed solution works
in the following procedure: The available sensor nodes are deployed
at initial locations to cover part of the RoI, and the collected data
is kept at the base station (throughout this paper, this data is called
\textit{historical data}). After some time (depending on the monitored
phenomenon's periodic behavior), some of the sensor nodes are moved
to other points that are not already covered in the initial deployment
(we call this as the \textit{moved subset}). At the same time, a subset
of the nodes are kept in their original locations (we call this as
the \textit{fixed subset}). This fixed subset is chosen by considering
the area spatial correlation characteristic and the required monitoring
accuracy, i.e., higher accuracy requires larger subset to be kept.
Thereupon, the historical data is used to train a supervised neural
network such that the input layer represents the fixed subset and
the output layer is for the moved subset. Therefore, at any time instance
and by using the fixed subset data, the uncovered locations (old locations
of the moved subset) can be reproduced.

Suppose that the collected historical data $T$ consists of $n$ samples
in the form $\mathbf{T}=\left\{ \left(\mathbf{\vec{x}}^{(1)},\mathbf{\vec{y}}^{(1)}\right),...,\left(\mathbf{\vec{x}}^{(n)},\mathbf{\vec{y}}^{(n)}\right)\right\} $,
where $\mathbf{\vec{x}}^{(i)}$ is the fixed subset data at time instance
$i$ and $\mathbf{\vec{y}}^{(i)}$ is the moved subset data at the
same instance. The predicted sensors' output $\mathbf{\vec{p}}^{(i)}$
is generated when the input vector $\mathbf{\vec{x}}^{(i)}$ is placed
at the neural network's input neurons. The historical samples are
used to train the neural network such as to minimize the sum of squares
of the error (SSE) between the original output and the network predicted
output as follows:
\[
\text{SSE}=\sum_{j=1}^{n}\left\Vert \vec{\mathbf{y}}^{(j)}-\mathbf{\vec{p}}^{(j)}\right\Vert ^{2}.
\]

Choosing the number of neurons in the input and the output layer is
a simple process. The input layer is formulated by a number of neurons
that is identical to number of fixed nodes. Similarly, the output
layer includes one neuron for each moved node. On the other hand,
choosing the size of the hidden layer of a neural network is a key
ingredient to achieve better estimation results. A widely accepted
design method is to choose the size of the hidden layer to be between
the sizes of the input and the output layers, i.e., to maintain the
pyramid shape of the neural network. Moreover, using more than one
hidden layer can efficiently enhance the neural network's estimation
ability. However, this increases the computation requirement of the
learning process, i.e., algorithm's execution time before convergence
to local minimum.

Moreover, we propose a hybrid mechanism to accelerate the convergence
of the backpropagation algorithm. Specifically, we noticed that the
Rprop method outperforms the BFGS algorithm with fewer iterations.
However, BFGS is more effective for minimizing the cost function over
long runs. Then, it is important to realize that each iteration of
BFGS requires the calculation of Hessian matrix. As a result, the
BFGS method is more complex than Rprop in terms of computational requirement.
A hybrid mechanism by starting the learning process using Rprop for
a few iterations, e.g., one-tenth the total learning iterations, and
then using BFGS for final tuning can significantly enhance the overall
process in terms of average error and learning time.

\section{Results and evaluation}

In this paper, we use a real data set from the Sensorscope project
\cite{barrenetxea2008wireless}. The data includes temperature readings
from a meteorological application. The sensors (23 sensors) sense
data in the range of -20 to 60 Celsius. To
evaluate the proposed solution, we consider the test case that only
14 sensors are available and mark the rest as uncovered locations.
Therefore, the 14 sensors are utilized to reproduce the data of the
23 sensors. Moreover, all the experiments in this paper use the cross-validation
method \cite{kohavi1995study} to test the solution efficiency. Accordingly,
the historical data set is divided into five groups. The training
is performed over four of them, while the remaining one is kept out
for testing purposes. In this way, the testing is performed using
data samples that are never seen by the generated model of the solution.

\begin{figure}
\begin{centering}
\includegraphics[width=0.75\columnwidth,trim=3cm 1.7cm 3cm 1cm]{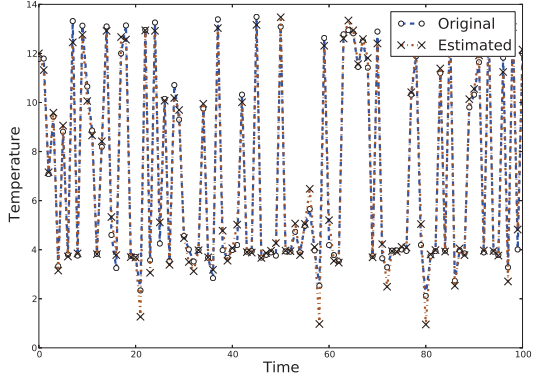}
\par\end{centering}

\caption{\label{fig:performance-analysis}Performance analysis: Original readings
and the algorithm estimation of them.}

\end{figure}

Figure \ref{fig:performance-analysis} contains data series of a sensor
node over time. Moreover, it includes estimated values assuming that
the sensor was moved from its location. Even though the monitored
area produces fluctuation pattern, the proposed method predicts reasonable
estimations when that location was uncovered by physical sensors.

\begin{figure}
\begin{centering}
\includegraphics[width=0.78\columnwidth,trim=3cm 1.5cm 3cm 1.5cm]{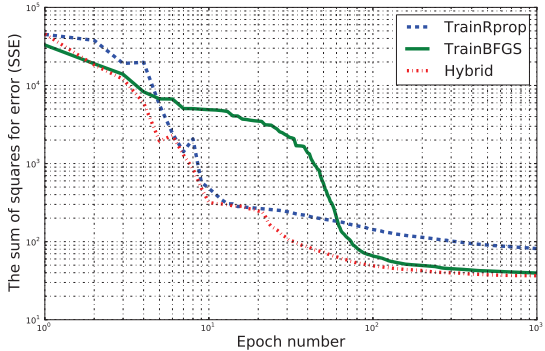}
\par\end{centering}

\caption{\label{fig:sse}Comparison between the backpropagation methods: Rprop,
BFGS, and the hybrid method.}
\end{figure}

The performance of Rprop, BFGS, and the hybrid method is presented
in Fig. \ref{fig:sse}. We use the sum of squares of error (SSE) to
quantize the difference between the original and the estimated values.
During an iteration of the learning process, the algorithm iterates
over all learning samples to tune the neural network parameters. In
Rprop, the weights are updated by multiplying the old value by a factor
$\alpha$. In particular, if there is no sign change between successive
iterations, the factor is set to be less than one, e.g., $\alpha=0.6$.
Otherwise, it is set to be greater than one, e.g., $\alpha=1.2$.
BFGS is a method that uses the Hessian matrix, i.e., a second order
method, to find the optimization direction of unconstrained and nonlinear
problems. BFGS converges to local minimum more accurately than the
Rprop method. On the negative side, BFGS takes more time to converge
than Rprop and requires more computational resources. Our hybrid method
benefits from both algorithm capabilities by starting the learning
process using Rprop, followed by BFGS at the second stage. This hybrid
technique significantly facilitates the convergence process to the
local minimum with a lower number of iterations.

For comparison purpose, the absolute error between the original and
the predicted outputs is defined as follows:
\[\varepsilon_{abs}=\frac{1}{n}\sum_{j=1}^{n}\left|\mathbf{\vec{y}}^{(j)}-\mathbf{\vec{p}}^{(j)}\right|.\]

Accordingly, we compare the estimation capabilities of the algorithms
with different number of hidden layers as summarized in Table \ref{tab:comparison_layers}.
Here, it is important to consider the tradeoff between the algorithm's
performance and the time required to learn the correlation among sensors
(the more layers, the lower the average error, and the more time required
to train the neural networks). However, if the network is not trained
over sufficient iterations, the algorithm performance will degrade
when using more layers, e.g., as in the case of using 5 layers instead
of 4 layers with BFGS (Table \ref{tab:comparison_layers}). 

\begin{table}
\centering{}\caption{\label{tab:comparison_layers}Comparison between the used neural network's
layers and methods over 1000 iterations. This includes the absolute
error $(\varepsilon_{abs})$ in Celsius and the required learning
time $\left(t_{learn}\right)$ in hours for each method. The learning
time is just used to compare between the three algorithms.}
\begin{tabular}{|>{\centering}m{2cm}||>{\centering}m{1.6cm}|>{\centering}m{1.6cm}|>{\centering}m{1.6cm}|}
\hline 
\textbf{\# of layers and neurons at each layer} & \textbf{Rprop}

$(\varepsilon_{abs},t_{learn})$ & \textbf{BFGS}

$(\varepsilon_{abs},t_{learn})$ & \textbf{Hybrid}

$(\varepsilon_{abs},t_{learn})$\tabularnewline
\hline 
\hline 
3 layers 

(14:11:9) & $(0.845,3.29)$ & $(0.639,4.99)$ & $(0.622,4.73)$\tabularnewline
\hline 
4 layers

(14:13:12:9) & $(0.820,5.01)$ & $(0.566,8.05)$ & $(0.554,7.40)$\tabularnewline
\hline 
5 layers (14:13:12:11:9) & $(0.812,6.57)$ & $(0.586,13.96)$ & $(0.536,12.66)$\tabularnewline
\hline 
\end{tabular}
\end{table}

\section{Summary and ongoing work}

In this paper, we have proposed a solution to the problem of monitoring
an area with a few number of sensor nodes. In particular, the method
explores the spatial correlation among sensor nodes using supervised
neural networks. Therefore, the proposed method predicts measurements
at uncovered zones. Moreover, we have shown that a hybrid method of
Rprop and BFGS can significantly enhance the learning stage in terms
of performance and execution time.

In ongoing research, we aim to develop a model to select the number
of Rprop and BFGS iterations in the hybrid method. Moreover, we will
analyze the statistical models of the area which helps in selecting
the moved and the static subsets of nodes, and we will connect this
with the error control and bounding.

\section*{Acknowledgment}

This work was supported by the A{*}STAR Computational Resource Centre
through the use of its high performance computing facilities.

\bibliographystyle{IEEEtran}
\bibliography{references}

\begin{thebibliography}{10}
\providecommand{\url}[1]{#1}
\csname url@samestyle\endcsname
\providecommand{\newblock}{\relax}
\providecommand{\bibinfo}[2]{#2}
\providecommand{\BIBentrySTDinterwordspacing}{\spaceskip=0pt\relax}
\providecommand{\BIBentryALTinterwordstretchfactor}{4}
\providecommand{\BIBentryALTinterwordspacing}{\spaceskip=\fontdimen2\font plus
\BIBentryALTinterwordstretchfactor\fontdimen3\font minus
  \fontdimen4\font\relax}
\providecommand{\BIBforeignlanguage}[2]{{%
\expandafter\ifx\csname l@#1\endcsname\relax
\typeout{** WARNING: IEEEtran.bst: No hyphenation pattern has been}%
\typeout{** loaded for the language `#1'. Using the pattern for}%
\typeout{** the default language instead.}%
\else
\language=\csname l@#1\endcsname
\fi
#2}}
\providecommand{\BIBdecl}{\relax}
\BIBdecl

\bibitem{batalin2002sensor}
M.~A. Batalin and G.~S. Sukhatme, ``Sensor coverage using mobile robots and
  stationary nodes,'' in \emph{ITCom 2002: The Convergence of Information
  Technologies and Communications}.\hskip 1em plus 0.5em minus 0.4em\relax
  International Society for Optics and Photonics, 2002, pp. 269--276.

\bibitem{costanzo2012nodes}
C.~Costanzo, V.~Loscr{\'\i}, E.~Natalizio, and T.~Razafindralambo, ``Nodes
  self-deployment for coverage maximization in mobile robot networks using an
  evolving neural network,'' \emph{Computer Communications}, vol.~35, no.~9,
  pp. 1047--1055, 2012.

\bibitem{liu2013dynamic}
B.~Liu, O.~Dousse, P.~Nain, and D.~Towsley, ``Dynamic coverage of mobile sensor
  networks,'' \emph{IEEE Transactions on Parallel and Distributed Systems},
  vol.~24, no.~2, pp. 301--311, 2013.

\bibitem{erdelj2013multiple}
M.~Erdelj, V.~Loscri, E.~Natalizio, and T.~Razafindralambo, ``Multiple point of
  interest discovery and coverage with mobile wireless sensors,'' \emph{Ad Hoc
  Networks}, vol.~11, no.~8, pp. 2288--2300, 2013.

\bibitem{gupta2008efficient}
H.~Gupta, V.~Navda, S.~Das, and V.~Chowdhary, ``Efficient gathering of
  correlated data in sensor networks,'' \emph{ACM Transactions on Sensor
  Networks (TOSN)}, vol.~4, no.~1, p.~4, 2008.

\bibitem{michaelides2011improved}
M.~Michaelides and C.~G. Panayiotou, ``Improved coverage in wsns by exploiting
  spatial correlation: the two sensor case,'' \emph{EURASIP Journal on Advances
  in Signal Processing}, vol. 2011, no.~1, pp. 1--11, 2011.

\bibitem{he2012leveraging}
S.~He, J.~Chen, X.~Li, X.~Shen, and Y.~Sun, ``Leveraging prediction to improve
  the coverage of wireless sensor networks,'' \emph{IEEE Transactions on
  Parallel and Distributed Systems}, vol.~23, no.~4, pp. 701--712, 2012.

\bibitem{rumelhart1988learning}
D.~E. Rumelhart, G.~E. Hinton, and R.~J. Williams, \emph{Learning
  representations by back-propagating errors}.\hskip 1em plus 0.5em minus
  0.4em\relax MIT Press, Cambridge, MA, USA, 1988.

\bibitem{riedmiller1992rprop}
M.~Riedmiller and H.~Braun, ``Rprop-a fast adaptive learning algorithm,'' in
  \emph{Proceedings of the International Symposium on Computer and Information
  Science}.\hskip 1em plus 0.5em minus 0.4em\relax Citeseer, 1992.

\bibitem{ngiam2011optimization}
J.~Ngiam, A.~Coates, A.~Lahiri, B.~Prochnow, Q.~V. Le, and A.~Y. Ng, ``On
  optimization methods for deep learning,'' in \emph{Proceedings of the 28th
  International Conference on Machine Learning}, 2011, pp. 265--272.

\bibitem{barrenetxea2008wireless}
G.~Barrenetxea, F.~Ingelrest, G.~Schaefer, and M.~Vetterli, ``Wireless sensor
  networks for environmental monitoring: the sensorscope experience,'' in
  \emph{IEEE International Zurich Seminar on Communications}.\hskip 1em plus
  0.5em minus 0.4em\relax IEEE, 2008, pp. 98--101.

\bibitem{kohavi1995study}
R.~Kohavi, ``A study of cross-validation and bootstrap for accuracy estimation
  and model selection,'' in \emph{International Joint Conference on Artificial
  Intelligence}, vol.~14, no.~2, 1995, pp. 1137--1145.

\end{thebibliography}

\end{document}